\documentstyle[aps,prl,epsf,rotate]{revtex}
\begin{document}

\twocolumn[\hsize\textwidth\columnwidth\hsize\csname     
@twocolumnfalse\endcsname

\title{Experimental Evidence for a Spin-Polarized Ground State in the 
$\nu=5/2$ Fractional Quantum Hall Effect
}

\author{W. Pan$^{1,2}$,
H.L. Stormer$^{3,4}$,
D.C. Tsui$^{1}$,
L.N. Pfeiffer$^{4}$, K.W. Baldwin$^{4}$, and
K.W. West$^{4}$}
\address{$^{1}$Department of Electrical Engineering, Princeton
University,
Princeton, New Jersey 08544}
\address{$^{2}$National High Magnetic Field Laboratory, 
Tallahassee, Florida 32310}
\address{$^{3}$Department of Physics
and Department of Applied Physics, Columbia University, New
York,
New York 10027}
\address{$^{4}$Bell Labs, Lucent Technologies, Murray Hill, New
Jersey 07974}

\date{\today}
\maketitle

\begin{abstract}

We study the $\nu=5/2$ even-denominator fractional quantum Hall effect 
(FQHE) over a wide range of magnetic field in a heterojunction 
insulated gate field-effect transistor (HIGFET). 
The electron density can be tuned from $n=0$ to $7.6 \times 10^{11}$ 
cm$^{-2}$ with a peak mobility $\mu = 5.5\times10^6$ cm$^2$/Vs. 
The $\nu=5/2$ state shows a strong minimum in diagonal resistance 
and a developing Hall plateau at magnetic fields ($B$) as high as 12.6T. 
The strength of the energy gap varies smoothly with $B$-field. 
We interpret these observations as strong evidence for 
a spin-polarized ground state at $\nu=5/2$.

\end{abstract}

\vskip2pc]

Ever since its discovery in 1987 \cite{willett:prl87} the origin of 
the even-denominator fractional quantum Hall 
effect (FQHE) state at Landau level filling factor 
$\nu= 5/2$ remains mysterious. Results of recent 
ultra-low temperature experiments \cite{pan:prl99} leave no 
doubt that this even-denominator state is a true 
FQHE with vanishing resistivity and Hall plateau formation. 
Such an even-denominator 
state does not fit the normal odd-denorminator rule of the FQHE 
and requires other or additional 
electron correlations. An early, so-called Haldane and Rezayi 
(HR) model \cite{hr:prl88} arrives at a {\em spin-unpolarized} trial 
wave function.

Eisenstein et al. \cite{eisenstein:prl88} tested the spin-polarization 
of the $\nu=5/2$ state by tilted magnetic 
field experiments \cite{fang:pr68} in a traditional, 
fixed-density sample. While the orbital motion 
of the electrons and hence their correlation 
energy ($E_c$) is subject only to the perpendicular 
component of the magnetic ($B$) field, the Zeeman energy ($E_z$) 
depends on the {\em total} $B$-field. Varying angle 
and $B$-field, the specimen can be kept in the 
$\nu=5/2$ state while the Zeeman energy is raised. 
Such a procedure should leave a spin-polarized state 
intact, but should be detrimental to a spin-unpolarized 
state, once the Zeeman energy cost surmounts the gain 
in correlation energy. In the experiment the strength 
of the $\nu = 5/2$ state decreased quickly upon tilting 
and the state disappeared totally at $\theta \sim 50^{\circ}$. 
This was taken as evidence of a spin-singlet state at $\nu=5/2$.

In recent years, with the advent of the 
composite fermion (CF) model \cite{jain,hlr}, there has 
been a renewed interest in the $\nu=5/2$ 
state \cite{mr:npb91,greiter:prl91,morf:prl98,park:prb98,fradkin:npb98,nick:prl99,ino:prl99,ardonne:prb00,jain:nature00,read:preprint00,yu:prb00,maeda:npb01}. 
Moore and Read (MR) proposed a ground state of 
$p$-wave paired CF's \cite{mr:npb91}. Unlike the HR state, 
the MR state is {\em spin-polarized}.
It is now being argued that the earlier disappearance 
of the $\nu=5/2$ state under tilt may be the result of 
the compression of the wave function due to the 
in-plane component of the $B$-field. This reduction of 
the $z$-extend of the wave function affects electron 
correlation and is the cause for the gap collapse, 
rather than the suspected increase in Zeeman energy. 
Indeed, recently two tilted field experiments \cite{pan:prl99a,lilly:prl99} 
showed that the added in-plane magnetic field 
not only destroys the FQHE at $\nu=5/2$ but also 
induces an electronic transport anisotropy. 
Theoretical modeling \cite{rezayi:prl00} suggests that this is 
due to a phase transition from the MR pairing 
state to an anisotropic state and unrelated 
to any spin-effect. Hence, the spin-polarization of 
the $\nu=5/2$ FQHE remains unresolved and there is 
presently little experimental input into the debate 
over the nature of the even-denominator FQHE state.

Here we pursue the spin-polarization of the $\nu=5/2$ 
state in analogy to the tilted field experiments 
by investigating the competition between
$E_c$ and $E_z$. Rather than 
tuning their ratio in a fixed density specimen 
by tilt, we keep the $B$-field perpendicular to the 
2DES and employ a variable density specimen. 
Since for a fixed filling factor, such as 
$\nu=5/2$, $E_c \propto n^{1/2}$, whereas $E_z \propto n$, 
increasing electron density modifies the ratio of 
$E_z$ to $E_c$. This approach is 
equivalent to tilting the sample, but it cannot cause 
a tilted-field induced phase transition \cite{note1}.
In our density-dependent experiment on the 
$\nu=5/2$ state we observe a strong minimum in 
diagonal resistance and a developing Hall plateau 
even at very high electron densities, equivalent to 
$B$-fields as high as 12.6T. Furthermore, the strength 
of the energy gap varies slowly and smoothly with $B$-field. 
We interpret these observations as evidence for a 
spin-polarized ground state at $\nu=5/2$.

To perform these experiments we fabricate a HIGFET 
(heterojunction insulated gate field-effect transistor), 
in which the 2DES density can be tuned from 
$n = 0$ to $7.6 \times 10^{11}$ cm$^{-2}$ with a peak 
mobility $\mu = 5.5 \times 10^6$ cm$^2$/Vs. The left inset of Fig.~1 
shows the profile of such a HIGFET \cite{kane:apl91}, 
with a heavily $n^+$-doped GaAs cap layer as the gate. 
In contrast to modulation-doped GaAs/AlGaAs 
heterostructures \cite{pfeiffer:apl89}, there is no 
delta-doping layer in a HIGFET, which allows to reach 
high mobilities at very high densities. We have 
investigated as many as 15 HIGFETs. In all specimens 
we observed qualitatively the same results. In this letter, 
we present experimental data from the specimen, 
which allowed for the highest gate-voltages and, 
hence, highest electron densities.

The right inset of Fig.~1 shows the strictly linear dependence of density 
$n = 1.542 \times V_g - 0.082$ on $V_g$ where $n$ is in units of 
10$^{11}$ cm$^{-2}$ and $V_g$ in units of volt. 
The mobility $\mu$, shown in Fig.~1, rises monotonically 
to a value as high as $\mu \sim 5.5 \times 10^6$ cm$^2$/Vs at 
$n = 5.5 \times 10^{11}$ cm$^{-2}$. Beyond this density 
electrons are starting to populate the second 
electrical subband \cite{stormer:ssc82}. This opens an additional 
scattering channels and leads to a reduction in mobility. 

We have measured $R_{xx}$ and $R_{xy}$ at $\nu = 5/2$ over 
the density range from $n = 3.0 \times 10^{11}$ to 
$7.6 \times 10^{11}$ cm$^{-2}$. The high-density limit is set 
by the breakdown voltage of the HIGFET and for densities 
much lower than $n = 3.0 \times 10^{11}$ cm$^{-2}$ the $\nu=5/2$ 
state becomes weak. In Fig.~2(a)-2(c), we show the diagonal 
resistance $R_{xx}$ in the interval $3 > \nu > 2$ at three 
selected densities, $n = (3.0, 5.3, 7.6) \times 10^{11}$ cm$^{-2}$. 
In all cases there exists a strong minimum in 
$R_{xx}$ at $\nu = 5/2$. This indicates the existence of a 
$\nu=5/2$ FQHE state over the whole range of density from $n=3.0$ 
to $n=7.6 \times 10^{11}$ cm$^{-2}$. The $\nu=5/2$ FQHE state 
at 12.6T in Fig.~2(c) represents, to our knowledge, 
by far the highest $B$-field at which this state has 
ever been observed. 

Since we are limited to standard 
dilution-refrigerator temperatures of $T>30mK$, 
$R_{xx}$ never vanishes in our HIGFET data, in contrast 
to our previous ultra-low temperature measurements \cite{pan:prl99}. 
For this reason we cannot determine the energy gap, $\Delta$, 
from activation energy measurement. In order to quantify our 
results we employ an earlier method \cite{gammel:prb88} 
to attribute a strength, $S$, to the $R_{xx}$ minimum. 
$S$ is defined as the ratio of the depth of the minimum 
to the average around $\nu=5/2$, $S= R_{5/2}/R_{ave}$ (see inset Fig. 3(a)).  
$S$ varies exponentially in temperature, 
$S \propto$ exp($-\Delta^{quasi}/2k_BT$) and defines a 
quasi-energy gap, $\Delta^{quasi}$. Since $S$ measures a quantity 
very similar to $R_{xx}$ and is proportional to it, 
at least at higher temperatures, we can take $\Delta^{quasi}$ 
to be very similar to $\Delta$.

Fig.~3(a) shows the strength, $S$, of the $\nu=5/2$ state versus 
inverse temperature (1/$T$) for three selected densities. 
For $n < 3.0 \times 10^{11}$ cm$^{-2}$ the $\nu=5/2$ becomes 
too weak to be quantifiable in terms of $S$. 
On this semilog plot the data follow a linear 
relationship and $\Delta^{quasi}$ is readily deduced. 
Fig.~3(b) shows the value of $\Delta^{quasi}$ at $\nu=5/2$ 
filling (solid dots) versus $B$-field. $\Delta^{quasi}$ 
hovers around 150mK and there is little variation 
over this magnetic field range.  Our previous, ultra-low 
temperature measurement \cite{pan:prl99} on a fixed 
density ($2.3 \times 10^{11}$ cm$^{-2}$) specimen yielded 
an activation energy of $\Delta$ = 110mK. A comparison between 
both data indicates that the value of the $\Delta^{quasi}$ 
is a good approximation to $\Delta$. 
In Fig.~3(b) $\Delta^{quasi}$ increases slightly from the 
smallest $B$-field to about 9T, whereupon it decays 
somewhat to higher fields. The dependence is smooth 
and no indication for any sharp transition is apparent.

From the data of Fig.~3(b) we conclude that it is highly unlikely 
that the $\nu=5/2$ state is spin-unpolarized. At 12.6T 
the Zeeman energy amounts to $E_z \sim$ 3K, which is more than 
a factor of 15 larger than the biggest energy gap ever 
measured for any $\nu=5/2$ state, which typically vary from 100-200mK. 
The overwhelming strength of the Zeeman energy should have 
overcome any correlation-induced spin-reversal and 
should have destroyed such a spin-unpolarized FQHE state. 
The smooth field-dependence of $\Delta^{quasi}$ in Fig.~3(b) 
also indicates that no phase transition occurs in the 
$\nu=5/2$ state over this field range. Therefore the 
polarization of the spin system remains unchanged, 
implying a {\em spin-polarized} state for the entire range of Fig.~3(b). 
For comparison we plot in the same figure the $\Delta^{quasi}$ of 
the $\nu=8/5$ state, measured in the same specimen.  
The well-studied transition from a spin-unpolarized 
to a spin-polarized state \cite{eisenstein:prl89,clark:prl89,furneaux:prl89,engel:prb92} expresses itself as a strongly reduced energy gap at $\sim$7.5T 
in contrast to the smooth variation of the gap at $\nu=5/2$.

The above qualitative argument can be quantified with the help of Fig.~4. 
Here we plot the dependence of $E_c=\alpha e^2/\epsilon l_B$ and 
$E_z=g\mu_BB$ versus $B$-field with $l_B = (\hbar c/eB)^{1/2}$ 
being the magnetic length, $\epsilon=12.9$ the dielectric 
constant of GaAs, and $\mu_B$ the Bohr magneton. We use $g=0.44$ 
for the $g$-factor in GaAs \cite{dobers:prb88} and 
$\alpha=0.02$ from a numerical 
calculation of Morf \cite{morf:prl98}. For a spin-unpolarized state 
one would take the energy gap to be $\Delta = E_c-E_z$ 
which is represented by the difference between $E_c$ 
and $E_z$ in Fig.~4. $E_z$ equals $E_c$ at a critical 
$B$-field of $B_{crit} \sim 11$T. For higher fields
the Zeeman energy exceeds the correlation energy and, 
in a simple model, one would expect a spin-unpolarized state 
to be no longer stable. With the usual 
uncertainties in the theory and the simple nature of our 
model 12.6T is insufficiently 
far from this $B_{crit} \sim$ 11T to totally rule out a spin-unpolarized 
ground state on this basis. However, theoretical energy 
gaps always exceed experimentally measured energy gaps. 
The reason is probably an inherent level broadening, $\Gamma$, 
due to disorder, which subtracts from the theoretical gap. 
From various experiments \cite{cfbook} this broadening is believed to 
be $\sim$0.5-1K for the standard high-mobility specimens and 
roughly magnetic-field independent within a given
specimen. Such a broadening shrinks the range of observable 
gaps as indicated in Fig.~4. We have chosen $\Gamma=0.6$K, 
which best reflects the value of 
$\Delta \sim$ 0.1-0.2K measured in most samples in the range of 3-5T. 
This reduces $B_{crit}$ to less than 7T, considerably 
lower than our highest value of 12.6T. Alternatively, 
in order to reflect the experimental values of 
$\Delta \sim$ 0.1-0.2K measured in most samples in the range of 
3-5T the theoretical value for $\alpha$ may be an overestimate 
and consequently $\Gamma$ may be smaller than $\Gamma \sim$ 0.6K. Yet, 
inspection of Fig.~4 shows that any such variation 
would only reduce the upper critical field. 
A theoretical value larger than $\alpha=0.02$ is very unlikely, 
contrary to the trends in the experimental data 
and contrary to the results of other few particle 
numerical calculations \cite{rezayi}. On the basis of these 
comparisons one can be very confident, that the 
$\nu=5/2$ FQHE state is {\it spin-polarized} over the 
entire density range over which it has been experimentally observed.

There remain some observations that 
need to be addressed. The gap of a spin-polarized 
system should follow the 
correlation energy $E_c \propto n^{1/2} \propto B^{1/2}$. 
The data of Fig.~3(b) initially show such a gradual 
increase but eventually turn around at $B \sim$ 9T. 
The origin of this behavior is unclear, but is 
probably related to the increasing confinement of 
the electrons against the interface due to the 
rising gate voltage. Electron scattering from 
residual interface roughness increases, leading to 
an increase in $\Gamma$ and a deceasing
energy gap for higher gate voltages and higher 
electron densities. (Note that such an $n$-dependent 
$\Gamma$ would further reduce $B_{crit}$ in Fig.~4). 
The decrease is not related to the population of 
a second subband as seen in the $B=0$ mobility data 
in Fig.~1, although both occur at similar densities.
In the 
range of the $\nu=5/2$ state Landau quantization is 
operative and a simple, $B=0$ intersubband scattering 
model is no longer valid. Furthermore, a comparison 
between Landau level splitting ($\sim$22meV) and electrical 
subband splitting ($\sim$19meV) at 12.6T indicates that 
the $\nu=5/2$ state resides in the second Landau 
level of the lowest electrical subband, as in all 
previously studied, fixed-density specimens. 
The absence of any discontinuity of $\Delta^{quasi}$ 
in Fig.~3(b) indicates that this condition 
persists over the whole density range investigated. 
In any case, the existence of a $\nu=5/2$ FQHE state, 
residing in the lowest Landau level of the second 
electrical subband is, {\it a priori}, very unlikely. 
Such a state is equivalent to $\nu=1/2$, where CF's 
form a fermi sea and not a CF-paired FQHE state.

In summary, in a variable-density specimen we 
observe a $\nu=5/2$ FQHE state at a magnetic field 
as high as 12.6T. The high Zeeman energy at this $B$-field 
and a detailed comparison with a universal model 
calculation exclude a spin-unpolarized ground state. 
The quasi-energy gap of the $\nu=5/2$ FQHE varies 
smoothly over the whole density range from 
$n=3.0$ to $7.6 \times 10^{11}$ cm$^{-2}$, which excludes a transition 
between spin-polarizations. From these findings we 
conclude that the even-denominator $\nu=5/2$ FQHE 
is spin-polarized, consistent with a Moore-Read paired CF state.

We would like to thank E. Palm and T. Murphy for experimental
assistance. A
portion of this work was performed at the National High Magnetic Field
Laboratory, which is supported by NSF Cooperative Agreement No.
DMR-9527035 and by the State of Florida. D.C.T. and W.P. are supported
by the AFOSR, the DOE, and the NSF.

\vspace*{-0cm}

\begin{figure}[t]
\epsfxsize=2.5in
\centerline{
\epsffile{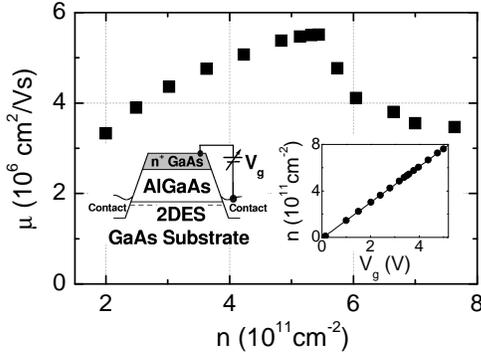}}
\caption{
Electron mobility as a function of density, $n$. The left inset depicts 
the layer structure of the HIGFET. The right inset shows $n$ 
versus $V_g$ for the HIGFET.
The solid circles ($\bullet$) are experimental data.
}
\end{figure}

\begin{figure}[t]
\epsfxsize=2.5in
\centerline{
\epsffile{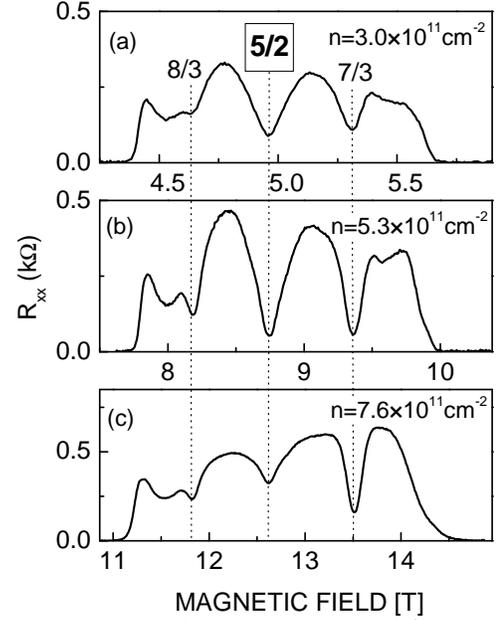}}
\caption{
Magneto-resistance around $\nu = 5/2$ at three densities at $T \sim$ 50mK. 
A low-frequency ($\sim$ 7Hz) lock-in technique with excitation 
current $I$ = 10 nA is used. The vertical lines mark the 
positions of the FQHE states at $\nu$ = 8/3, 5/2, and 7/3.
}
\end{figure}

\begin{figure}[t]
\epsfxsize=2.5in
\centerline{
\epsffile{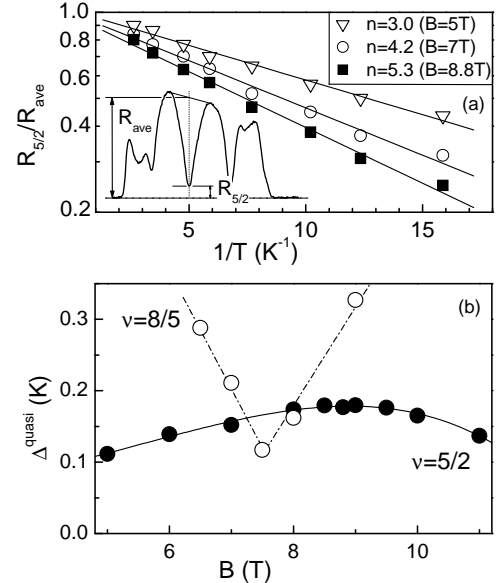}}
\caption{
(a) Arrhenius plot for $R_{5/2}/R_{ave}$ at three densities, 
in units of 10$^{11}$ cm$^{-2}$. (b) ($\bullet$) Smooth variation of 
quasi-energy gap of the $\nu$ = 5/2 FQHE state as a function 
of magnetic field ({\it i.e.} electron-density). 
($\circ$) Collapse of the $\nu=8/5$ quasi-energy gap due to 
the well-documented transition in its spin-polarizations.
}
\end{figure}

\begin{figure}[t]
\epsfxsize=2.5in
\centerline{
\epsffile{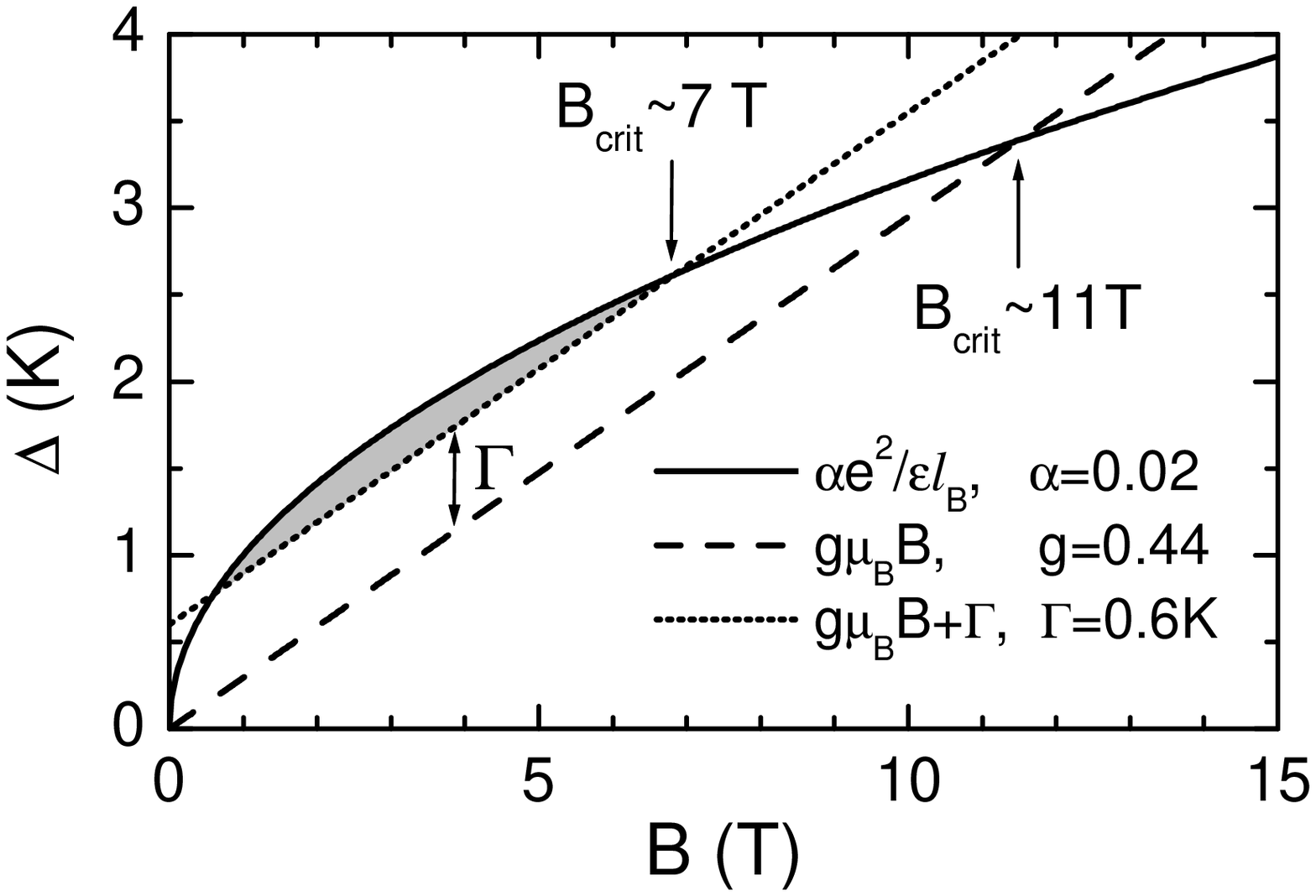}}
\caption{
Model calculation of the energy gap at $\nu=5/2$ as a 
function of magnetic field. The solid line is the 
Coulomb energy $e^2/\epsilon l_B$, the dashed line is 
the Zeeman energy $g \mu_B B$, and the dotted line is 
the sum of Zeeman energy and disorder broadening, $\Gamma$, 
of CF's. For discussion see text.
}
\end{figure}


\begin{references}
\bibitem{note1}
The $z$-extent of the wavefunction also varies in the
HIGFET as $<z> \propto n^{-1/3}$.
However, since $l_B \propto n^{-1/2}$ also changes,
the parameter $<z>/l_B \propto n^{1/6}$,
which is relevant for electron-electron correlation,
has only a very weak density dependence.
\end{references}
\end{document}